\documentclass[fleqn,usenatbib]{mnras}
\usepackage{amsmath}
\usepackage[T1]{fontenc}
\usepackage[utf8]{inputenc}
\usepackage[pdftex]{graphicx}
\usepackage{amsmath}
\usepackage{float}
\usepackage{xcolor}
\usepackage{amssymb}
\usepackage{newtxtext,newtxmath}

\title[The HI-halo mass relation at $z \sim 1$]{The HI-halo mass relation at redshift $z \sim 1$ from the Minkowski functionals of 21~cm intensity maps}

\author[Spina et al.]{Benedetta Spina$^{1,}$\thanks{E-mail: bspina@astro.uni-bonn.de}, Cristiano Porciani$^{1}$ \& Carlo Schimd$^{2}$ \\
$^{1}${Argelander Institut f\"ur Astronomie , Auf dem H\"ugel 71, 53121 Bonn, Germany,}\\
$^{2}${Aix Marseille Univ, CNRS, CNES, LAM, Marseille, France}
}

\date{Accepted XXX. Received YYY; in original form ZZZ}

\pubyear{}

\begin{document}
\label{firstpage}
\pagerange{\pageref{firstpage}--\pageref{lastpage}}
\maketitle

\begin{abstract}
The mean and the scatter of the HI content of a dark-matter halo as a function of the halo mass are useful statistics that can be used to test models of structure and galaxy formation. 
We investigate the possibility of constraining this HI-halo mass relation (HIHMR) from intensity maps of the redshifted
21~cm line.
In particular, we use the geometry and topology of the brightness-temperature isocontours in a single frequency channel as quantified by the Minkowski functionals.
First, we generate mock maps from a large N-body simulation considering the impact of thermal noise and foreground removal.
We then use the Fisher information formalism to forecast constraints on a parametric model for the HIHMR.
We consider a 20,000 deg$^2$ survey (originally proposed for dark-energy science) conducted with the
Square Kilometre Array Phase 1 (SKA-1) MID observatory operating in single-dish mode. For a channel bandwidth of 2 MHz, we show that an integration time of a few$\,\times\,10^4$ s per pointing is sufficient to image the smoothed HI distribution at redshift $z\simeq 1$ and to measure the HIHMR in a nearly optimal way from the Minkowski functionals. Tighter constraints on some of the parameters can be obtained by using also an independent measurement of the mean HI density. 
Combining the results from different frequency channels 
provides exquisite constraints on the evolution of the HIHMR, especially in the central frequency range of the data cube.
\end{abstract}



\section{Introduction}

Cold-gas reservoirs in galaxies provide the raw fuel for star formation. Assessing how they vary across different galaxy populations and environments is of paramount importance to constrain models of galaxy assembly and evolution.

Neutral hydrogen (HI) makes up the bulk of the cold gas.
In the post-reionisation Universe, HI can be found almost exclusively
within self-shielded clouds inside galaxies and galaxy clusters.
The total HI content of dark-matter (DM) haloes is thus
a simple descriptive statistic that can be used to compare
theoretical models with observations.
It gives the HI mass in a halo by
summing up the contributions from central and satellite galaxies (as well as from diffuse gas). 
A strong theoretical prejudice is that this quantity should mainly
depend on the halo mass. In fact, gas-accretion rates 
(both from the intergalactic medium and via galaxy mergers)
are expected to be regulated by the halo masses and so are also
several processes that deplete the HI content (like the efficiency of galactic winds or photoionisation).
Extra dependencies on top of the halo mass will appear as scatter
around the mean relation. Measuring the amplitude of this scatter provides an empirical test of the assumption that halo mass is the main driver of the HI content. 

The HI-halo mass relation (HIHMR)
has been the subject of many studies
based on (post-processed) hydrodynamical simulations 
\citep{Dave+13, Villaescusa+14, Crain+17, 21cmIM, Ando+19} and
semi-analytical models of galaxy formation \citep{Kim+17,Zoldan+17,baugh2019,spinelli2019}.
In parallel,
observational constraints at redshift $2.2 \lesssim z \lesssim 5$ have been obtained for damped Lyman-$\alpha$ absorbers \citep{Barnes-Haehnelt14}. On the other hand,
in the local Universe, extensive investigations have been carried out
to elucidate the HIHMR of
HI-rich galaxies from the Arecibo Fast Legacy ALFA Survey
\citep[ALFALFA,][]{Giovanelli+05}.
For instance, 
\citet{Paul+18} have predicted the HIHMR
by assuming that there exists a scaling relation
between the HI content and the optical properties of a galaxy.
By cross matching the ALFALFA HI sources with the optical group catalog from the Sloan Digital Sky Survey, \citet{ai-zhu-18} have estimated the total HI mass for rich
galaxy groups containing eight members or more.
Combining a similar technique with information about the HI-weighed clustering of the ALFALFA sources, 
\citet{Obuljen+2019} have constrained the parameters of an analytical model for the HIHMR.
A direct measurement of the relation has been recently performed by
\citet{guo+20} who stacked the ALFALFA spectra of the members of the
SDSS groups.

At intermediate redshifts ($0.1\lesssim z\lesssim 2.2$),
we have very little information on the HI content of DM haloes 
as the 21~cm line is too faint to observe individual galaxies with
reasonable integration times.
A way out is to
detect the collective emission from galaxies that occupy large `voxels' in our past-light cone defined by the angular resolution and the bandwidth of the observations (21~cm tomography).
The principal challenge for these intensity-mapping experiments is the subtraction of
Galactic and extra-galactic foregrounds that are orders of magnitude brighter than the 21~cm signal.
In order to filter them out, their properties 
and the instrumental systematics need to be characterized to extremely precise levels.

There is increasing interest in developing large-scale surveys that 
map the intensity of the 21~cm emission from neutral hydrogen 
to probe unprecedented volumes.
Such experiments are expected to provide pivotal contributions to cosmology \citep[e.g.][]{BarkanaLoeb2005,Loeb-Wyithe08,Monsalve+2019} and, at higher redshifts, to our understanding of the epoch of reionisation \citep[e.g.][]{YueFerrara2019}.

Information about
the HIHMR is key for 
predicting the signal of 
21~cm intensity-mapping experiments. 
In full analogy with the halo model 
for dark-matter \citep[e.g.][]{Cooray-Sheth02} and galaxy clustering \citep[e.g.][] {Scoccimarro+01}, the HI distribution on large scales can be described in terms of its halo-occupation properties.
By using the halo model to fit 
a compilation of data at low and high redshift,
\citet{Padmanabhan+15} have estimated that
the amplitude of the intensity-mapping signal 
at intermediate redshifts is between 50 and 100 per cent uncertain.
\citet{Switzer+13} have obtained
the first observational constraint on the amplitude of 21~cm fluctuations at $z \simeq 0.8$.
There are some difficulties, however, in fitting this result together with the low-redshift studies and the Damped Lyman-$\alpha$ systems (DLAs) data into a consistent picture
\citep{Padmanabhan-Refregier17, Padmanabhan2017}.

In this work, we take the opposite step and investigate the possibility of inferring the HIHMR from the 21~cm intensity maps.
However, in order not to weaken the constraining power of these experiments for cosmology, we do not use the power spectrum and focus on the geometrical and topological properties of the highly non-Gaussian maps as quantified by the Minkowski functionals (MFs) of their isocontours.
Several morphological indicators  
have been already discussed in the 21~cm literature
to characterise the growth of HII bubbles during the epoch of reionisation, namely, the genus curve \citep{Hong+14}, the MFs \citep{GleserNusserCiardiDesjacques2006,Chen+2019}, and the Minkowski tensor \citep{Kapahtia+2018}. 
Here, we apply one of these methods to the post-reionisation Universe in order to constrain a parametric model for the HIHMR (mean and scatter). We use mock data based on a large N-body simulation and the Fisher information matrix to derive the constraints on the model parameters. 
The rationale behind our method is rather simple: since DM haloes of different masses trace the underlying mass-density distribution differently, the detailed morphology of the brightness-temperature maps should reflect the HIHMR. 

We analyse here two-dimensional maps corresponding to individual
frequency channels in a tomographic data cube.
The main reason for this choice is that we can robustly estimate
the covariance matrix of the data without having to run hundreds
of high-resolution N-body simulations. Our method, however, can be straightforwardly generalised to three-dimensions by directly measuring the MFs in the full data cube.

The proposed approach requires rather long radio integration times as it is necessary to image the redshifted 21~cm signal 
with a sufficient signal-to-noise ratio to measure
the MFs reliably. In order to reduce the observing time, 
we smooth the maps with an isotropic Gaussian filter before measuring the MFs.

We compute how the constraints on the HIHMR vary 
with the observing time for 
an array of radio telescopes used in single-dish mode.
As a reference case, we consider
the Square Kilometre Array Phase 1 (SKA-1) MID observatory.
We provide a proof of concept of the method
by only considering mock data at $z\simeq 1$
but, of course, there is no particular difficulty to repeat the investigation for other redshifts.

The paper is organised as follows. 
In section~\ref{sec:21cm}, we introduce the HIHMR and explain how we use it to generate mock intensity maps starting from a large N-body simulation. The definitions of the MFs and their application to the 21~cm intensity maps are presented in section~\ref{sec:MF}. Our implementation 
of the Fisher matrix formalism is described in section~\ref{sec:fisher}.
Our results are presented in section~\ref{sec:results} and discussed
in section~\ref{sec:discussion}.
Finally, in section~\ref{sec:conclusion}, we briefly summarize the philosophy behind our approach and our main findings.

\section{Mock 21~cm intensity maps}\label{sec:21cm}
In the post-reionisation Universe, HI is mostly found within
the DM haloes that host galaxies and galaxy clusters.
However, it is a matter of fact that
HI intensity-mapping experiments will survey large fractions of the sky.
Therefore, in order to forecast the constraining power of the forthcoming 21~cm experiments for cosmology and astrophysics, we need to simulate the large-scale structure of the Universe over sizeable volumes while resolving low-mass structures and accounting for the physics that regulates hydrodynamics, feedback, and radiative transfer in the interstellar medium. Fulfilling all these requirements together is prohibitive for state-of-the-art software and computing facilities.
A possible way forward is to operate at a simpler level of understanding by combining high-resolution N-body simulations with a statistical description of the HI content of DM haloes. 
In this section,
we describe how we construct mock HI intensity maps following this approach.
Finally, we illustrate how we account for the different sources of noise in the intensity maps.


\subsection{N-body simulation}

We use the \textsc{MultiDark-Planck (MDPL)} simulation \citep{BolshoiSimulations} 
that assumes a $\Lambda$CDM background and considers a set of cosmological parameters which is compatible with the fit in \citet{Planck13}. 
The simulation evolves $N=3840^3$ DM particles of mass $m_{\mathrm p} = 1.51 \times 10^9 h^{-1}\mathrm{M}_\odot$ 
(where the Hubble constant is written as $H_0=100\,h$ km s$^{-1}$ Mpc$^{-1}$)
in a (periodic) cubic box of comoving side $L = 1\, h^{-1}\, \rm{Gpc}$.
DM haloes are identified using a standard `friends-of-friends' algorithm with a linking length of $\ell=0.2\,L/N^{1/3}$.
Only haloes containing at least 20 particles are considered in this work.


\subsection{HI-halo mass relation}
\label{sec:relation}
A common working hypothesis is that mass is the main driver
for the HI content of a DM halo.
In this case, it makes sense to introduce the HIHMR
$M_\mathrm{HI}(M)$
that gives the mean HI mass found in a DM halo of mass $M$
\citep{Pontzen+08, bagla2010, Barnes-Haehnelt14, Seehars+16, Padmanabhan-Refregier17, Padmanabhan2017, 21cmIM,baugh2019}.
We adopt the following functional form,
\begin{equation}
    M_\mathrm{HI}(M) = M_0 \left( \frac{M}{M_\mathrm{min}} \right)^\alpha \exp \left[-\left( \dfrac{M_\mathrm{min}}{M} \right)^{\gamma} \right]\;.
\label{eqn:MHI_Mh}
\end{equation}
The HIHMR is a scale-invariant power law of slope $\alpha>0$
with an exponential cutoff at small halo masses
(i.e. for $M\lesssim M_\mathrm{min})$.
This suppression reflects the fact that low-mass haloes cannot self-shield from the UV background and gas cooling is inhibited in them \citep[e.g.][]{Rees86, Efstathiou92}.
The parameter
$\gamma>0$ determines how sharp the cutoff is while $M_0$ fixes the overall normalisation of the HIHMR -- note that $M_\mathrm{HI}(M_\mathrm{min})=M_0/e$. 
For our forecasts at $z=1$, we assume the fiducial values
$M_0=1.5\times10^{10}h^{-1}M_\odot$, $M_\mathrm{min} = 6.0\times10^{11}h^{-1}\mathrm{M}_\odot$, $\alpha=0.53$, and
$\gamma=0.35$ which provide an accurate description of the Illustris TNG simulation \citep[][where the local HI density is obtained by subtracting the molecular fraction estimated with a chemical-equilibrium model from the total hydrogen abundance]{21cmIM} 
and are also in agreement with the current observational estimates for the HI density \citep[see e.g.][]{crighton2015,hu2019}.

In reality, many other factors influence
the HI content of a DM halo, for instance, hydrodynamic processes as well as radiative and mechanical feedback from star formation and accretion onto compact objects. 
The clustering properties of HI-rich galaxies in the present-day Universe suggest that halo spin also plays a role in determining the cold-gas content \citep{Papastergis+13}.
We treat these secondary dependencies beyond halo mass in a statistical way as scatter around the HIHMR.
Therefore, we assume that, at fixed halo mass, $M_\mathrm{HI}$ 
is a random variable that follows a lognormal distribution whose mean is given by equation~(\ref{eqn:MHI_Mh}).
This is equivalent to say that $\ln [M_\mathrm{HI}/(1\,h^{-1}M_\odot)]$ (at fixed $M$) is a Gaussian random variable with mean $\ln [M_\mathrm{HI}(M)/(1\,h^{-1}M_\odot)]-\sigma^2/2$ and
standard deviation $\sigma$.
We use the fiducial value of $\sigma=1$ as it approximately matches the scatter measured in the
semi-analytical models analysed in \citet{spinelli2019}.


\subsection{From haloes to brightness temperature} \label{sec:haloes-temperature}

We associate a HI mass to each DM halo by randomly sampling the corresponding lognormal distribution. 
We then use the Cloud-In-Cell mass-assignment scheme to 
build a map of the HI density on a regular Cartesian grid with $210^3$ cells.
In order to account for redshift-space distortions, we deposit the HI
at the location
\begin{equation}
    \mathbf{s} = \mathbf{x} + \frac{1+z}{H(z)} \mathbf{v}_\parallel,
\end{equation}
where $\mathbf{x}$ is the actual comoving position of the halo,
$\mathbf{v}_\parallel$ is its peculiar velocity along the line of sight, and $H(z)$ denotes the Hubble parameter in the background.
Note that this neglects both the relative position and the relative motion of the neutral hydrogen with respect to the halo centre of mass.

Eventually, we compute the brightness temperature using
\begin{equation}
\delta T_b (\mathbf{s}) = 189\,h\, \dfrac{H_0}{H(z)}\, (1+z)^2\, \dfrac{\rho_{\rm HI}(\mathbf{s})}{\rho_{\mathrm{c},0}}\, \mathrm{mK}\;,
\label{eqn:Tb-HI}
\end{equation}
where $\rho_{\mathrm{c},0}$ is the critical density of the Universe at redshift $z=0$ and $H_0$ denotes the Hubble constant.


\subsection{Frequency bandwidth and angular resolution}
\label{sec:angular_resolution}
\begin{figure*}
\includegraphics[width=1.8\columnwidth]{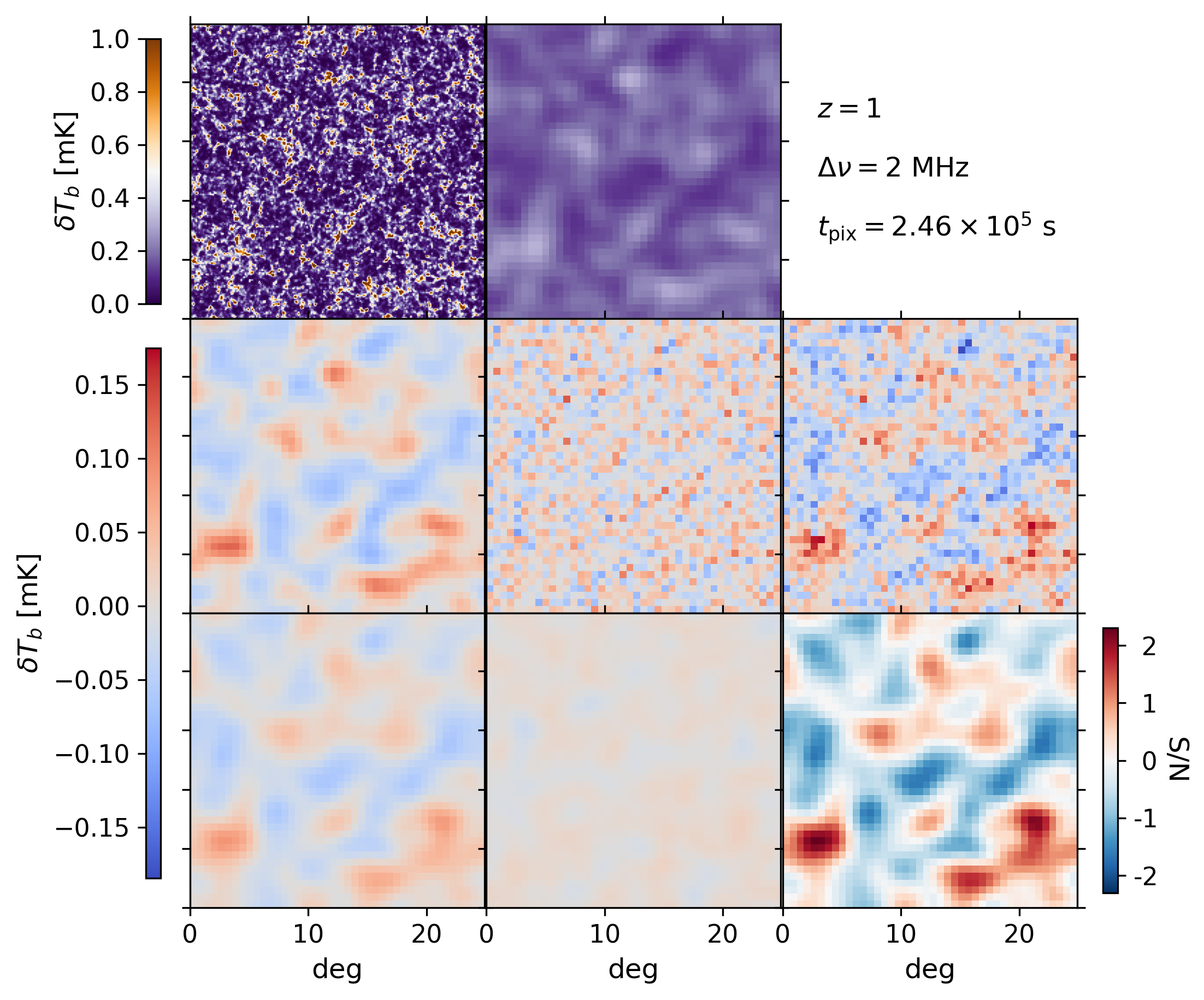}
\caption{Steps towards building mock 21~cm intensity maps. 
The original brightness-temperature distribution associated with the DM haloes in the simulation (top-left panel, see section~\ref{sec:haloes-temperature}) is smoothed to simulate
the finite angular resolution of radio telescopes (top-right, see section~\ref{sec:angular_resolution}). Long-wavelength Fourier modes in the data cube are lost during the foreground removal (section~\ref{sec:foregrounds})
which shifts the mean signal towards zero (centre-left, note the different color scale).
Thermal noise is generated at the pixel level (central panel, see section~\ref{sec:instnoise}) and
added to the signal (centre-right). 
Finally, some smoothing is applied as a data processing technique to enhance the signal-to-noise ratio (bottom-left). The smoothed noise map and
the signal-to-noise map are shown in the bottom-central and bottom-right panels, respectively.
}
\label{fig:IM}
\end{figure*}

We now perform a mock 21~cm tomography of the simulated data.
For each pointing of the radio telescope, 
spectroscopic information is collected by partitioning
the total receiver bandwidth into a number of frequency channels.
The signal-to-noise ratio of radio images depends critically on 
the channel bandwidth (see section~\ref{sec:instnoise}).
By indicating with $\chi$ the radial comoving distance in the background cosmological model, a channel bandwidth $\Delta \nu$ at redshift $z$ corresponds to the radial separation
\begin{equation}
\Delta \chi
= \int_{z_{-}}^{z_{+}} c \dfrac{\mathrm{d} z}{H(z)}
\end{equation}
where 
\begin{equation}
z_{\mp}= \dfrac{1}{\displaystyle{\frac{1}{1+z}\pm\frac{\Delta \nu}{2\nu_{\mathrm{rest}}}}}-1\;.
\label{eqn:z+-}
\end{equation}
For a rest-frame frequency of $\nu_{\mathrm{rest}}=1420.406$ MHz,
the cell size of the Cartesian grid we use to sample the HI density
corresponds\footnote{We adopt here the distant-observer (plane-parallel) approximation and use one of the axes of the simulation box as the line of sight.} to $\Delta \nu\simeq 1$ MHz at $z=1$. In order to produce synthetic maps with a larger $\Delta \nu$,
we average $\delta T_b$ at a fixed position on the sky over the corresponding length scale $\Delta \chi$.
This way, we can produce $210\,(1\, \mathrm{MHz}/\Delta \nu)$ non-overlapping intensity maps of the 21~cm signal at $z=1$.
One example with $\Delta \nu=2$ MHz is shown in the top-left panel of
figure~\ref{fig:IM}. Wherever some HI is found along the line of sight,
a brightness temperature fluctuation of a few mK is recorded.

In order to account for the finite angular resolution of the instrument, we convolve the two-dimensional maps with the telescope beam.
For a single dish of diameter $D$,
we assume a Gaussian beam with full width at half maximum\footnote{The coefficient of $1.2$ represents a correction for the non-uniform illumination of the antenna and is not related to the Rayleigh criterion. In the ideal case, the numerical factor would be $1.02$ \citep[e.g.][]{hall2005,chen2020}.}  
\begin{equation}
\theta_{\rm FWHM}=  1.2\, \dfrac{\lambda_{\rm rest}\, (1+z)}{D},
\label{eqn:theta_FWHM}
\end{equation}
and perform the convolution in Fourier space.
After setting $D = 15\, \mathrm{m}$ (for SKA-1 MID) and $\lambda_{\mathrm{ rest}} = 21.16$ cm,
this corresponds to an isotropic Gaussian smoothing in the plane
of the sky with standard deviation
\begin{equation}
\Sigma= \dfrac{\theta_{\rm FWHM}\, (1+z)\,d_{\mathrm{a}}}{2\sqrt{2\ln 2}}\;,
\label{eqn:Sigma}
\end{equation}
where $d_{\mathrm{a}}$ is the angular-diameter distance to redshift $z$ in the background (in a flat universe, $(1+z)\,d_{\mathrm{a}}=\chi$). It follows that, at $z = 1$, $\theta_{\rm FWHM} \approx 1.94\, \rm{deg}$ and $\Sigma \approx 33.1\, h^{-1}\,\rm{Mpc}$. 
Once convolved with the beam, the typical 21~cm signal from $z=1$
assumes values of the order of 0.1 mK (see the top-right panel of figure~\ref{fig:IM} for an example).


\subsection{Thermal noise}
\label{sec:instnoise}
The output of a radio telescope is contaminated by thermal noise. 
For single-dish observations, the noise is Gaussian to good approximation. 
Based on the radiometer equation, the rms noise fluctuation associated
with an integration time (per pointing) $t_\mathrm{pix}$ is 
\begin{equation}
    \sigma_\mathrm{N} = \frac{T_\mathrm{sys}}{\sqrt{t_\mathrm{pix}\,\Delta\nu}}
    \label{eqn:noise}
\end{equation}
where $T_\mathrm{sys}$ is the total system temperature.
For an antenna of SKA-1 MID, the system temperature can be obtained by summing up several components \citep{SKAredbook}
\begin{equation}
    T_{\rm sys}(\nu) = T_{\rm spl} + T_{\rm CMB} + T_{\rm gal}(\nu) + T_{\rm rx}(\nu),
\end{equation}
where $T_{\rm spl} \approx 3\, \rm{K}$ 
and $T_{\rm CMB} \approx 2.73 \, \rm{K}$ denote
the spill-over and the cosmic-microwave-background contributions, respectively.
The Galaxy contribution can be modelled as
\begin{equation}
    T_{\rm gal}(\nu) = 25 \left(\dfrac{408\, \rm{MHz}}{\nu} \right)^{2.75}\, \rm{K}
\end{equation}
and the receiver-noise temperature as
\begin{equation}
    T_{\rm rx}(\nu) = 15\,\rm{K} + 30 \left(\dfrac{\nu}{\rm{GHz}} - 0.75 \right)^{2}\, \rm{K}.
\end{equation}
For the 21~cm line emitted at $z = 1$, we obtain $T_\mathrm{sys} = 26.22$ K. The rms noise is thus much larger than $\delta T_b$
and integration times of a few days per pointing are needed to see the HI signal emerge above the noise at the pixel level (for a narrow bandwidth of $\Delta \nu=2$ MHz). 

We now imagine to build a pointed map.  
Assuming that multiple antennas can be used simultaneously,
the total time required 
to map the large-scale structure of the Universe 
over the solid angle $\Omega_\mathrm{surv}$
is then
\begin{equation}
    t_\mathrm{obs} = \frac{t_\mathrm{pix}}{N_\mathrm{dish}\, N_\mathrm{beam}} \frac{\Omega_\mathrm{surv}}{\Omega_\mathrm{pix}}\;,
    \label{eq:ttot}
\end{equation}
where $N_\mathrm{dish}$ is the number of available dishes (each with $N_\mathrm{beam}$ feedhorns, typically 1 or 2) and $\Omega_\mathrm{pix}  \simeq \theta_\mathrm{pix}^2 $ is the solid angle covered by a single pixel. 
In order to satisfy the Nyquist-Shannon theorem and produce
a properly sampled map, 
$\theta_\mathrm{pix}$ must be smaller than $\theta_{\mathrm{FWHM}}/2$.
In practice, it is usually chosen to be between $\theta_{\mathrm{FWHM}}/7$ and and $\theta_{\mathrm{FWHM}}/3$
\citep{marr2015fundamentals}. 

As an example, we estimate the time that would be needed to map the HI distribution in the MDPL box at $z=1$. The simulation box subtends a solid angle $\Omega_\mathrm{surv}\simeq 591$ deg$^2$ and we use $42^2$ pixels to cover the whole area. Assuming to observe with all the 197 antennas of the SKA-I MID telescope, we find $t_{\mathrm {obs}}\simeq 9 \,t_{\mathrm {pix}}$ (for $N_{\mathrm{beam}}=1$), i.e. of the order of a month for $\Delta \nu=2$ MHz.

\subsection{Foreground removal} \label{sec:foregrounds}
Reaching the necessary sensitivity is only one of the difficulties that we need to face in order to map the large-scale structure of the Universe in 21~cm.
Actually, the major challenge is the presence of Galactic and extra-galactic foregrounds 
that are several orders of magnitude brighter than the HI signal.
In particular, extragalactic point sources and
synchrotron emission from the Milky Way give major contributions. 
At first, it might seem impossible to separate these components.  
However, foregrounds should vary smoothly as a function of frequency (at a given position on the sky) while the fluctuating
21~cm signal is expected to be poorly correlated in frequency space. 
It has been suggested that this difference could be exploited
to separate the 21 cm signal from the foregrounds
\citep[e.g.][]{Shaver+99, Dimatteo+02, Oh-Mack03, Zaldarriaga+04}.
Although the concept is very promising, complex details need to be taken into account in practical implementations  like, for instance, dealing with frequency-dependent beam shapes.

In spite of the difficulties,
several foreground-cleaning techniques have been proposed and tested against simulated data \citep[see e.g.][for a recent review]{ska-Wolz}.
In addition to separating the foregrounds, these methods
usually introduce unintended consequences such as removing large-scale power from the 21~cm signal.
Following \citet{alonso2017} and \citet{cunnington2019}, 
we assume that the foreground subtraction erases all the Fourier modes
of $\delta T_b$
with comoving radial wavenumber 
\begin{equation}
    k_\parallel < k_\parallel^{\rm FG} \approx \dfrac{\pi \,H(z)}{c\,(1+z)\,\xi}\;,
\end{equation}
where $\xi$ is a dimensionless parameter of order unity.
Basically, $k_\parallel^{\rm FG}$
denotes the minimum wavenumber for which foregrounds are separable from the signal. It is difficult to assign a precise value to $\xi$ (which is method dependent), however, assuming $\xi \simeq 0.1$ appears to be a reasonable estimate \citep{cunnington2019}. 
For the cut-off scale at $z=1$, we therefore adopt
$k_\parallel^{\rm FG} \approx 0.01\, h\,\mathrm{Mpc}^{-1}$.
This value is only slightly larger than the fundamental frequency $k_{\mathrm{f}}=2\pi/L$ associated with our simulation box.
We thus only erase the Fourier modes with
$k_\parallel=0$ and $k_\parallel=k_{\mathrm {f}}$.
Note that subtracting the $\mathbf{k}=\mathbf{0}$ mode 
has the important consequence of setting the mean value
of $\delta T_b$ (over the whole simulation box) to zero (from the original value of approximately $0.2$ mK in the beam-convolved maps, cf. the top-right and centre-left panels in figure~\ref{fig:IM}). This, of course, makes the sensitivity
requirement for detecting the signal even more demanding.


\subsection{Smoothed maps}
The middle-left panel in figure~\ref{fig:IM} shows the 21~cm signal that one would obtain after the foreground removal but in the absence of thermal noise. 
In this particular map, $\delta T_b$ has a mean value of -0.0065 mK and an rms scatter of 0.035 mK. 
The adjacent panel on the right-hand side
shows a realisation of the thermal noise (assuming $t_\mathrm{pix} = 2.46\times 10^5$ s) with an rms scatter of 0.038 mK.
Comparing at face value these two maps 
gives the wrong impression that 68 hours of integration time
might not be enough to image the line emission coming from $z=1$.
This is because the 21~cm signal has been smoothed by the telescope beam while we have generated the thermal noise at the pixel level
(as we are assuming to have a different pointing per pixel). 
The observed map (i.e. the sum of the signal and the noise) is shown in the middle-right panel of figure~\ref{fig:IM}.
In order to reduce the impact of the fine-grained noise on the measurement of the MFs, we further smooth the observed map
with a two-dimensional Gaussian filter that has the same width as the telescope beam
(we adopt Gaussian smoothing for simplicity although Wiener filtering would be a nearly optimal choice given that the signal is only weakly non-Gaussian, see section~\ref{sec:discussion}).
In the bottom panels, we show, from left to right, the smoothed total map, the smoothed noise map (in which the rms scatter drops to 0.007 mK), and the corresponding signal-to-noise ratio (which assumes the peak value of 2.3).
It is evident that the smoothed observed map (bottom-left) presents the same characteristic features of the original one (middle-left) and can be used  
to study the morphological properties of the 21~cm signal.
This is the subject of the next section. 

\begin{figure*}
\includegraphics[width=2\columnwidth]{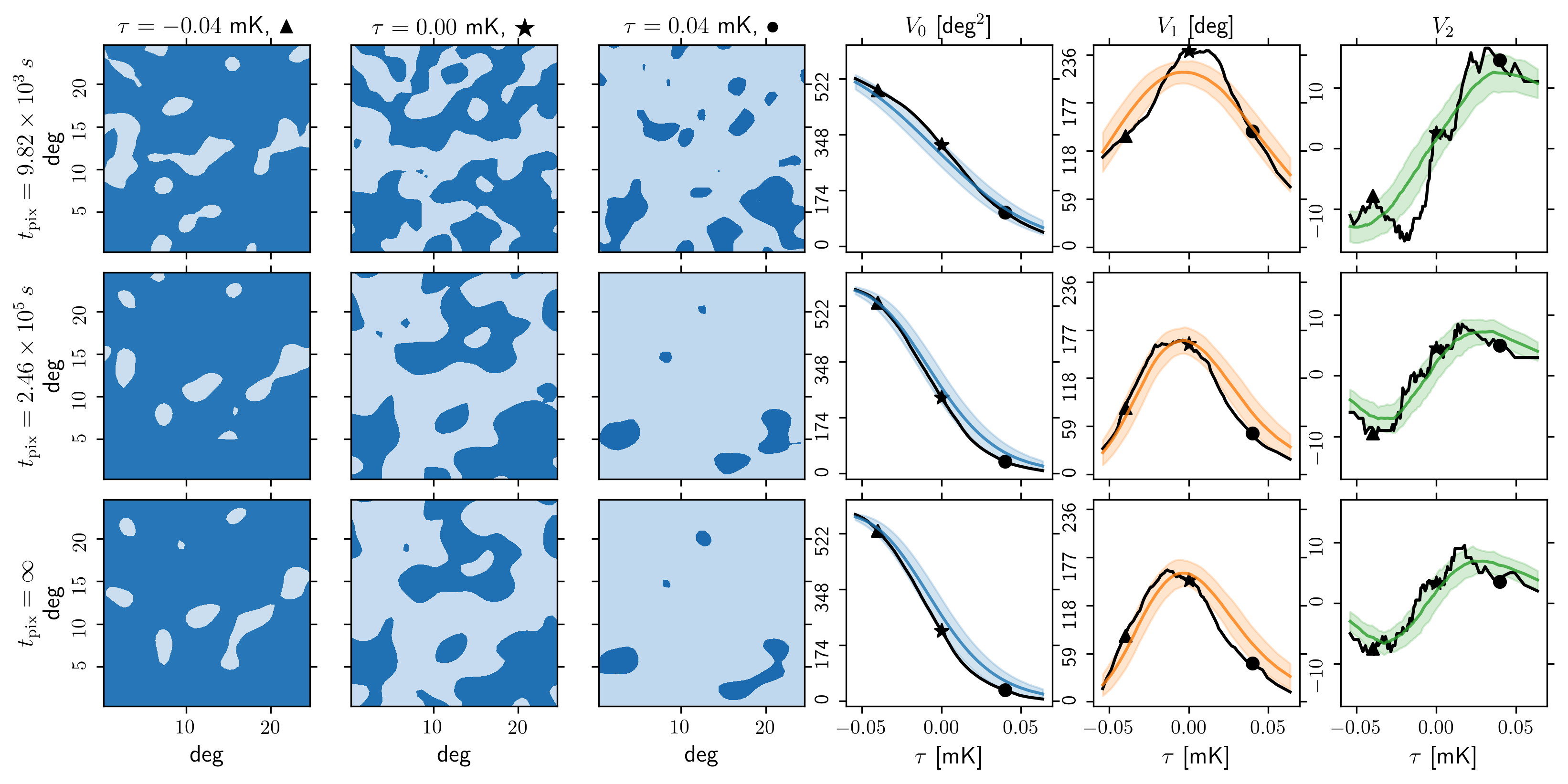}
\caption{In order to investigate the impact of thermal noise on the MFs, we consider here three intensity maps obtained at $z=1$
(with $\Delta\nu = 2$ MHz)
from the same region of the sky presented in figure~\ref{fig:IM}
but obtained varying the integration time per pixel (from top to bottom, $t_\mathrm{pix}=9.82\times10^3$ s, $2.46\times10^5$ s, $\infty$). The left set of figures shows the
regions above (dark blue) and below (light blue) 
three brightness-temperature thresholds (from left to right, $\tau = -0.04$ mK, 0 mK, 0.04 mK).
The right set of figures shows the corresponding MFs 
for the selected threshold values (triangle, star, and circle symbols)
and as a function of $\tau$ (black solid lines).
As a reference, we also show the mean (coloured solid lines) and the rms scatter (shaded regions) of the MFs extracted from the 105 volume slices (with transverse size $L^2$ and line-of-sight thickness corresponding to $\Delta \nu=2$ MHz) that fill the MDPL simulation box at $z=1$.
}
\label{fig:IM_MF}
\end{figure*}

\begin{figure}
\includegraphics[width=1\columnwidth]{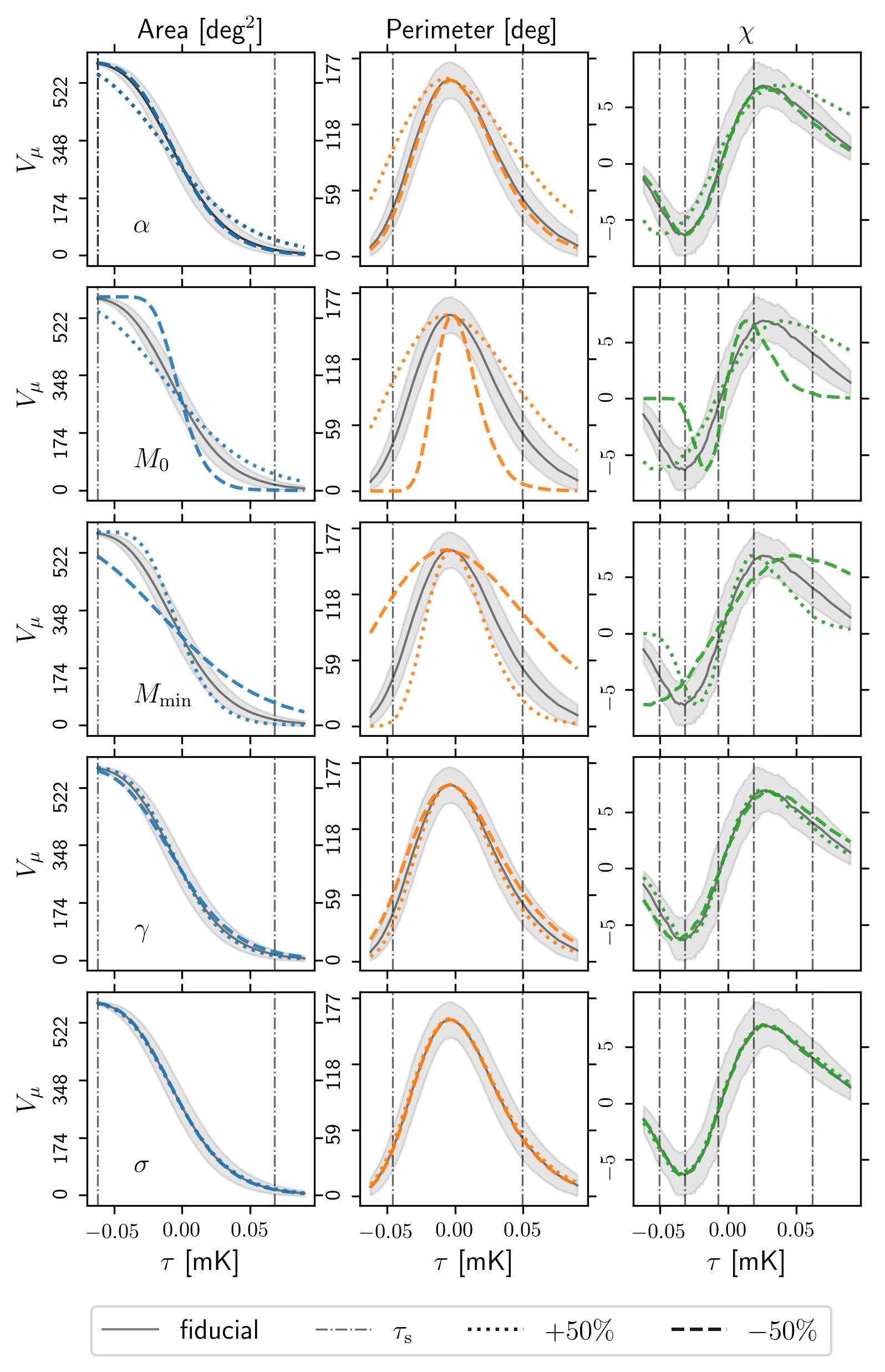} 
\caption{Dependence of the MFs on the parameters of the HIHMR in the absence of thermal noise (i.e. assuming $t_\mathrm{pix}\to \infty$). We vary each parameter by $\pm 50$ per cent with respect to the fiducial value at $z=1$. Shown are the corresponding
MFs averaged over the 105 volume slices that cover the MDPL simulation box (dotted and dashed lines). As a reference, we also show the results for the fiducial case (black solid lines) and their rms scatter (shaded region). The vertical dash-dotted lines indicate the set of thresholds we use in the final analysis presented in Section~\ref{sec:fisher}. We make use of slightly different values when we account for thermal noise.
}
\label{fig:MFvar}
\end{figure}

\section{Minkowski Functionals} \label{sec:MF}

\subsection{Basics}
A digital image can be thought of as a polyconvex set 
formed by the finite union of compact and convex subsets in $\mathbb{R}^d$ (its pixels). For such a system, the field of integral geometry provides a family of morphological 
descriptors known as MFs (or quermassintegrals or intrinsic volumes).
Basically, these functionals \citep{Minkowski1903}
measure the size and the connectivity of subsets of $\mathbb{R}^d$ in terms of different quantities.
In $d$ dimensions, there exist $d+1$ MFs that we denote with the symbols $V_0, \dots, V_d$ (as frequently done in the mathematical literature for the intrinsic volumes). 
Hadwiger's completeness (or characterisation) theorem \citep{hadwiger1957} states that the linear combinations of the MFs,
${\mathcal F}=\sum_{i} f_i \,V_i$ (where $f_i \in \mathbb{R}$ and the index $i$ runs from 0 to $d$), are the only functionals that satisfy the following properties:
i) additivity, ${\mathcal F}(A \cup B) ={\mathcal F}(A)+{\mathcal F}(B)
-{\mathcal F}(A \cap B)$;
ii) invariance under rigid motion, ${\mathcal F}(gA)={\mathcal F}(A)$
where $g$ is an element of the group of translations and rotations in $\mathbb{R}^d$;
iii) conditional continuity under the Hausdorff measure.
In this sense, the MFs completely characterize the morphology of a compact subset of $\mathbb{R}^d$.
All the notions above can be generalised to non-Euclidean spaces
of constant curvature, like the celestial sphere \citep[see e.g.][and references therein]{SchmalzingGorski1998}.

The MFs have been first introduced in cosmology by \citet{mecke1994} and \citet[][see also \citet{Gott1990}]{SchmalzingBuchert1997} in order to characterize the large-scale distribution of galaxies.
Subsequently, they have been used to measure the morphology of 
temperature anisotropies in the cosmic microwave background
\citep[e.g.][]{SchmalzingGorski1998, Novikov-Feldman-Shandarin-99}.
More recently, MFs have been employed to study the morphology and characterize the different stages of cosmic reionisation \citep{GleserNusserCiardiDesjacques2006,Chen+2019}.

\subsection{Application to 21~cm maps} \label{sec:application}
Although it would be possible to compute the MFs of $\delta T_b$ 
in the three-dimensional data cube, we perform here a simpler experiment and compute the MFs of two-dimensional maps corresponding to one frequency channel. The main motivation for this choice is that we can more easily estimate the covariance matrix for the measurements without having to generate hundreds of high-resolution simulation boxes (see also section~\ref{sec:fisher}).
To this end, we do not build a mock light cone by stitching together different snapshots of the MDPL simulation but only analyse the simulation output at $z=1$. As already mentioned in section~\ref{sec:angular_resolution}, this provides us with
$210\,(1\, \mathrm{MHz}/\Delta \nu)$ non-overlapping intensity maps of the 21~cm signal at $z=1$. 
Although these maps are not fully independent 
(due to correlations along the line of sight), we use them to derive the expected
signal in one channel (i.e. the average over the different maps) and the corresponding covariance matrix (see section~\ref{sec:fisher} for further details).

Let us consider the brightness-temperature map corresponding to a frequency channel and chose a threshold value $\tau$. The excursion set $E_{\tau}=\left\lbrace \boldsymbol{\varphi}\in\Omega: \delta T_{b}(\boldsymbol{\varphi}) > \tau \right\rbrace$ collects all the
points where $\delta T_b$ exceeds the threshold.
We define the MFs of the excursion set as

\begin{gather}
    V_0 (\tau) =  \int_{E_{\tau}} \!\!\!\!\!\mathrm{d}\Omega\;, \\
    V_1 (\tau) = \int_{\partial E_{\tau}} \!\!\!\!\!\!\!\mathrm{d}\ell\;, \\
    V_2 (\tau) = \frac{1}{2\pi} 
     \int_{\partial E_{\tau}} \!\!\!\!\!\!\! \kappa \,\mathrm{d}\ell\;,
\end{gather}
where $\mathrm{d}\Omega$ is the surface element of $E_{\tau}$,  $\mathrm{d}\ell$ is the length element of its (smooth) boundary $\partial E_{\tau}=\left\lbrace \boldsymbol{\varphi}\in\Omega: \delta T_{b}(\boldsymbol{\varphi}) = \tau \right\rbrace$, and $\kappa$ is the local geodesic curvature of $\partial E_{\tau}$.
In simple words, $V_0$ gives the 
surface area covered by the excursion set while $V_1$ is 
the perimeter of its boundary. Finally, in the flat-sky approximation used here, $V_2$ coincides with the Euler characteristic $\chi$ of $E_{\tau}$,
i.e. the number of connected regions in the excursion set, $N_+$, minus the number of holes, $N_-$.

There exist efficient algorithms to compute the MFs for digitised images 
\citep[e.g.][]{SchmalzingBuchert1997,SchmalzingGorski1998}. 
We use the public code
\textsc{minkfncts2d}\footnote{https://github.com/moutazhaq/minkfncts2d}
which only exploits information within the image and does not require boundary conditions\footnote{Although our synthetic maps are extracted from simulations with periodic boundary conditions, we analyse them as one would do for observational data.} as described in \citet{mantz2008}.
Isolines are built with the `marching square' algorithm which 
uses contouring cells obtained by combining
$2\times 2$ blocks of pixels. In consequence, given an input image with $N^2$ pixels, the code analyses the central $(N-1)^2$ pixels and neglects a narrow `frame' with half-pixel thickness lying along the boundary (corresponding to an area of $2N-1$ pixels).

Some illustrative examples are provided in 
the left half of figure~\ref{fig:IM_MF}. We use
the slice with $\Delta \nu=2$ MHz presented in figure~\ref{fig:IM}, observe it for three different
observation times $t_\mathrm{pix}$ (rows, increasing from top to bottom), and consider three
different temperature thresholds (columns, increasing from left to right). The dark regions in each sub-panel represent the excursion set (above threshold). 
The right half of the figure shows the MFs as a function of the
threshold temperature. The triangle, star and circle highlight the values for the images displayed in the panels on the left. The black solid line, instead, shows how the MFs change when many more thresholds are considered. This trend is compared with the mean
and the standard error obtained by averaging over the 105 slices
contained in our simulation box (coloured lines and shaded
regions).
The usefulness of the MFs can be more clearly understood by connecting the different plots. 
When $\tau$ is low (left column),
the whole observed domain is included in the 
excursion set with the exception of a few isolated holes.
Therefore, $V_0$ is large while $V_1$ is small 
(as only the boundaries of the holes contribute to it)
and $\chi$ is negative as in `swiss-cheese' sets dominated by the holes.
On the other hand, when $\tau$ approaches the median value of the
brightness temperature (middle panels), $V_0$ assumes intermediate values, $V_1$ reaches a local maximum, and $\chi$ is close to zero as in `sponge-like' structures where the regions above and below threshold are interconnected.
Finally, when $\tau$ is large (right panels), only a few isolated regions form the
excursion set. In this case, both $V_0$ and $V_1$ are low while 
$\chi$ is large and positive as in meatball topologies dominated by isolated connected regions.
Although this trend holds true for every smooth temperature distribution, the morphological properties of $\delta T_b$ 
are encoded in the precise shape of the $V_i(\tau)$ curves.

It is interesting to investigate
how the general pattern is altered by thermal noise.
As $t_\mathrm{pix}$ is reduced (moving from the bottom up), more and more small-scale structures appear in the excursion sets that substantially modify the values of the MFs.
In particular, their presence reduces the range of variability of $V_0$, shifts
$V_1$ towards larger values, and increases the extreme positive and negative values of $V_2$.

The sensitivity of the MFs to the HIHMR is investigated in figure~\ref{fig:MFvar}, in which we individually vary the parameters of the relation given in equation~(\ref{eqn:MHI_Mh}) by $\pm 50$ per cent and compare the results to the fiducial case and its standard error. Notice that each parameter modifies differently,
and to a varying extent, the shape of the MFs and the location of their extremal points.
In particular, increasing $M_0$ has a similar effect as decreasing $M_\mathrm{min}$ and vice versa. The figure clearly shows that information on the HIHMR is encoded
in the MFs.
Modifying the parameters of the HIHMR also changes
the mean HI density of the Universe and, through equation~(\ref{eqn:Tb-HI}),
the amplitude of $\delta T_b$. However, this systematic shift is lost due to the procedure of foreground subtraction. 

The bottom panel of figure~\ref{fig:MFvar} shows that the MFs are basically unaffected by relatively large variations of the parameter $\sigma$. This reflects the fact that the
comoving volume associated with a cylinder of radius $\Sigma$ and 
height $\Delta \nu$ is large enough to wash out 
any stochasticity in the HIHMR. For this reason, we do not consider $\sigma$ any further in this work
and only provide forecasts for the mean HI mass as a function of halo mass.

\subsection{Selected observables} \label{sec:obs}
In practice, we need to select a finite number of threshold values
with which to perform the measurements.
Although, in principle, one might want to consider a large number of thresholds, there is an actual limitation due to the fact that we need to estimate the covariance matrix of the measurement errors.
Since we can only use $S=105$ slices for this purpose, we need to limit the size of the data set.
The smoothness of the black solid curves in figure~\ref{fig:IM_MF} reveals that the MFs measured from the same map using
different values of $\tau$ are strongly correlated. This effect is particularly evident for $V_0$ and $V_1$.
After long experimenting in the attempt to minimize these correlations (and get
nearly optimal constraints on the HIHMR using a small number of thresholds), we end up picking two values of 
$\tau$ for $V_0$, two for $V_1$ and five for $V_2$ (as indicated in figure~\ref{fig:MFvar}).

In order to simplify the notation, we
combine all the measurements into a single 
seven-dimensional data vector $\mathbf{M}$. 

The parameters in 
equation~(\ref{eqn:MHI_Mh}) also determine the mean HI density, $\bar{\rho}_\mathrm{HI}(z)$, and thus the overall level of
the fluctuations, $\delta T_b$.
However, due to the foreground subtraction, we cannot measure
the mean brightness temperature of the 21~cm fluctuations.
This fact might weaken the constraints that the MFs can impose on the HIHMR. In fact, the differences between the curves
in figure~\ref{fig:MFvar} would be larger if one could measure the shift in brightness temperature associated with the change in the mean HI density.
In order to account for this missing piece of information, we combine the measurements of the MFs with observational constraints
on the cosmic abundance of HI, conventionally parameterized as $\Omega_\mathrm{HI}(z)=\bar{\rho}_\mathrm{HI}(z)/\rho_{\mathrm{c},0}$ \citep[for a recent review see e.g.][]{Peroux-Howk20}.
Current estimates at $z\sim 1$ are not fully consistent.
In fact, while
some studies of DLAs at $z\sim 1$ find $\Omega_\mathrm{HI}\simeq (6\pm 2)\times 10^{-4}$ \citep{Rao+06, Rao+17}, other estimates are approximately a factor of three lower \citep{Neeleman+16}.
Similarly, stacking the 21~cm emission from star-forming galaxies gives $\Omega_\mathrm{HI}<3.7 \times 10^{-4}$ \citep{Kanekar+16}.
Attempts to combine data from low to high redshifts and fit
the evolution of the HI abundance with a smooth curve give  $\Omega_\mathrm{HI}(z=1)= (6.1\pm 0.4)\times 10^{-4}$ \citep{crighton2015, hu2019}.
We use this result to set constraints on the parameters of the HIHMR.

\section{Fisher matrix}
\label{sec:fisher}
\subsection{Minkowski functionals} \label{sec:fisher_minkowski}
Let us now imagine to fit the measurements of the MFs with a theoretical model $\mathbf{M}_\mathrm{mod}(\boldsymbol{\theta})$ that depends on a set of tunable parameters, $\boldsymbol{\theta}$.
Assuming that the measurement errors follow a multivariate Gaussian distribution, we can write 
the likelihood function $\mathcal{L}$ of the model parameters as
\begin{equation}
    -2 \ln \mathcal{L}(\boldsymbol{\theta}) \propto (\mathbf{M}- \mathbf{M}_\mathrm{mod})^T \cdot\mathsf{C}^{-1} \cdot(\mathbf{M} - \mathbf{M}_\mathrm{mod})\;,
\end{equation}
where $\mathsf{C}$ denotes the covariance matrix of the data.
In order to forecast the model constraints that can be set by future measurements, we use the Fisher-information formalism.
We thus compute the Fisher matrix with elements
\begin{equation}
    F_{ij} = \dfrac{\partial\, \mathbf{M}_\mathrm{mod}^T}{\partial \theta_i} \cdot \mathsf{C}^{-1}\cdot \dfrac{\partial\, \mathbf{M}_\mathrm{mod}}{\partial \theta_j}\;.
    \label{eqn:fisher_elements}
\end{equation}
This expression assumes that $\mathsf{C}$ is computed for the fiducial model and not varied with the model parameters (as routinely done in cosmological studies).
We finally use $\mathsf{F}^{-1}$ as a proxy for
the asymptotic covariance matrix of the estimates for $\boldsymbol{\theta}$. This matrix represents the main result of our
study.

It is not easy to build a theoretical model for the MFs of the
21~cm signal.
For a zero-mean Gaussian random field $y$ over a two-dimensional space, the MFs can be expressed in terms of the power spectrum \citep{Adler81, Tomita90, NovikovSchmalzingMukhanov2000}, 
\begin{gather}
    V_0 (\tau) = \dfrac{1}{2} \mathrm{erfc} \left( \dfrac{\tau}{\sqrt{2}\sigma_0} \right), \label{eqn:V0} \\
    V_1 (\tau) = \dfrac{1}{2^{7/2}} \dfrac{\sigma_1}{\sigma_0} \exp \left[ -\dfrac{1}{2} \left(\dfrac{\tau}{\sigma_0}\right)^2 \right], \label{eqn:V1} \\
    V_2 (\tau) = \dfrac{1}{2^{5/2}\pi^{3/2}} \left( \dfrac{\sigma_1}{\sigma_0} \right)^2 \dfrac{\tau}{\sigma_0} \exp \left[ -\dfrac{1}{2} \left(\dfrac{\tau}{\sigma_0}\right)^2 \right], \label{eqn:V2}
\end{gather}
where $\tau$ is the threshold defining the excursion set, see Section~\ref{sec:application},
$\sigma_0^2=\langle y^2\rangle$ is the variance of the Gaussian random field, and $\sigma_1^2 = \langle y_{;i}y^{;i}\rangle$ (where the index $i$ runs from 1 to 2 and we adopted Einstein summation convention) is the variance of its covariant derivative.
For a weakly non-Gaussian field, they can be perturbatively expanded in terms of `skewness parameters' \citep{matsubara2003}. However, in the general case, one must rely on numerical simulations.
A convenient method for 21~cm tomography is to paint HI on top
of DM haloes extracted from a N-body simulation as we have done
to generate our mock data cubes. 
In this case, at fixed cosmology,
the model parameters coincide with those of the HIHMR.
Therefore, we replace $\mathbf{M}_\mathrm{mod}(\boldsymbol{\theta})$ with the expectation
value of our mock observations $\bar{\mathbf{M}}(\boldsymbol{\theta})$, i.e. averaged over sample variance and thermal noise. 

\begin{figure}
\includegraphics[width=\columnwidth]{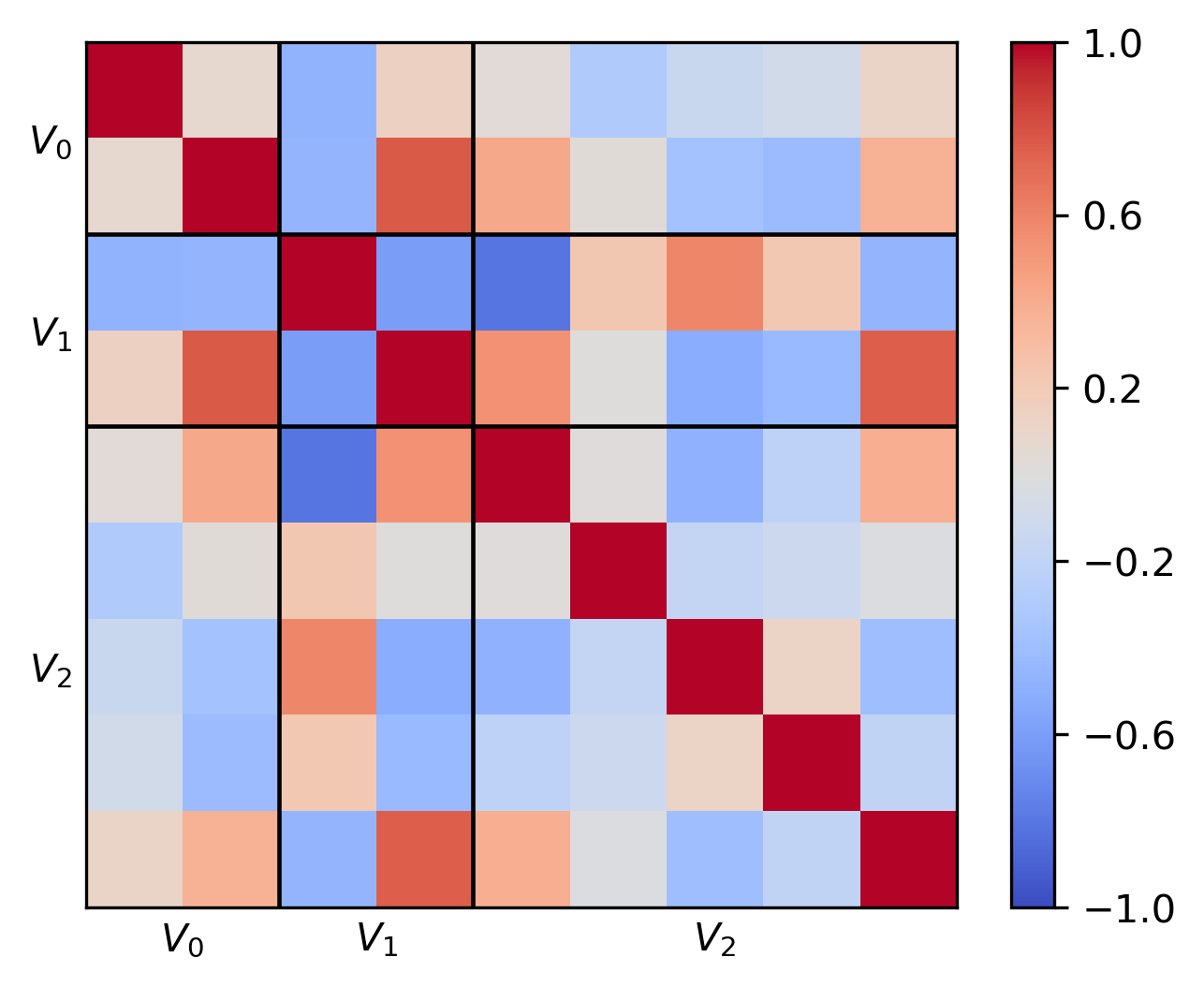}
\caption{The sample correlation matrix of the MFs for $t_\mathrm{pix} \rightarrow \infty$.
}

\label{fig:correlation}
\end{figure}

We first consider the limit
$t_\mathrm{pix} \to \infty$, in which thermal noise is irrelevant.
In this case, we denote the MFs obtained
from the $s^\mathrm{th}$ slice 
with the symbol $M_i^{(s)}$ where the index $i$ runs from 1 to 9.
We compute the expectation values of the MFs as 
\begin{equation}
    \bar{M}_i (\boldsymbol{\theta})= \dfrac{1}{S} \sum_{s=1}^S  M_i^{(s)}(\boldsymbol{\theta})\;.
\label{eqn:meanVnonoise}
\end{equation}
These quantities represent the mean measurement that would be  
obtained by averaging over the sample variance and 
many realisations of thermal noise. Of course, this is an abstract
quantity which is useful to compute the Fisher matrix but 
cannot be measured in practice as we can only access one
data set. 
Similarly, we approximate $\mathsf{C}$ with the sample covariance
matrix of the $S$ slices:
\begin{equation}
    \hat{C}_{ij} = \dfrac{1}{S-1}\,\sum_{s=1}^S \left( M_i^{(s)} - \bar{M}_i \right)\, \left( M_j^{(s)} - \bar{M}_j\right)\;,
\label{eqn:covariance-tpixinf}
\end{equation}
where all quantities are evaluated using the fiducial set of model parameters.
The corresponding correlation matrix is shown in figure~\ref{fig:correlation}. Particularly strong correlations are
noticeable in the $V_1$ sector and among the first and last thresholds for $V_2$. Note that, by using equation~(\ref{eqn:covariance-tpixinf}), we are implicitly assuming that the measurements extracted from neighbouring slices are independent while there are certainly some correlations along the line of sight.
Anyway,
we have checked that considering only one slice every two or three
does not change the overall structure of $\hat{\mathsf{C}}$.
Equation~(\ref{eqn:covariance-tpixinf}) gives an unbiased (but noisy, due to the fact that $S$ is finite) estimate of the covariance matrix. However, since matrix inversion is a highly non-linear operation, the resulting $\hat{\mathsf{C}}^{-1}$
is a biased estimate of the precision matrix.
In order to (statistically) correct for this bias, 
we compute the Fisher matrix by replacing ${\mathsf{C}}^{-1}$ with
$\Gamma\,\hat{\mathsf{C}}^{-1}$ where $\Gamma=\frac{S-13-2}{S-1}\simeq 0.865$ \citep{kaufman67,hartlap2007}.

We make sure that the numerical evaluation of the partial derivatives in equation~(\ref{eqn:fisher_elements}) gives stable results when the increment of the model parameters is changed. Before differentiating, we preventively smooth the function $V_2(\tau)$ using a Savitzky-Golay filter that irons out small-scale fluctuations appearing due to the fact that the Euler characteristics is integer valued (see the right panels in figure~\ref{fig:IM_MF}). This is a necessary step that makes the derivatives and the Fisher matrix meaningful. Our results are stable with respect to reasonable changes in the free parameters of the Savitzky-Golay method.

Considering the impact of thermal noise adds an extra level of complication. 
In order to extract the expected signal as a function of $t_\mathrm{pix}$ we proceed as follows.
We consider 10 different values for $t_\mathrm{pix}$ and, for each of them, we generate $N=30$ different realisations of the thermal noise over the full MDPL box. 
We then combine them with the 21~cm signal obtained using a particular
set of parameters $\boldsymbol{\theta}$ and
measure the MFs in each slice (frequency channel) of the foreground-cleaned data cubes. We denote the MFs obtained
from the $s^\mathrm{th}$ slice and the $n^\mathrm{th}$ noise realisation
with the symbol $M_i^{(s,n)}$. 
We compute the expectation values of the MFs as 
\begin{equation}
    \bar{M}_i (\boldsymbol{\theta})= \dfrac{1}{S\times N} \sum_{s=1}^S \sum_{n=1}^N M_i^{(s,n)}(\boldsymbol{\theta})\;,
\label{eqn:meanV}
\end{equation}
and substitute them for $\mathbf{M}_\mathrm{mod}(\boldsymbol{\theta})$ in equation~(\ref{eqn:fisher_elements}).
However, a problem arises with the estimate of the covariance matrix. 
We find, in fact, that $\hat{\mathsf{C}}$ depends on the noise realisation, in particular when $t_\mathrm{pix}$ becomes small. This has important consequences as the calculation
of the Fisher matrix requires inverting $\Gamma^{-1}\,\hat{\mathsf{C}}$
which prevents us from averaging the estimates of the covariance matrix
over the noise realisations.
We thus compute a Fisher matrix and produce forecasts for each noise realisation independently. We present our results in terms of the
distribution of the parameter constraints.

\subsection{Scaling with the survey area}
So far, we have always considered a region of the sky
with a linear transverse size that matches the comoving side of the MDPL simulation box. By considering only half or a quarter of this region,
we have checked that the data covariance matrix defined in equation \eqref{eqn:covariance-tpixinf} scales proportionally
to the survey area. Therefore, in order to consider a larger survey area, it is enough to correct the covariance matrix by the multiplicative factor $\zeta=\sqrt{\Omega_\mathrm{surv}/\Omega_\mathrm{MDPL}}$. 

\subsection{Constraints from $\Omega_\mathrm{HI}$} \label{sec:constraintsOHI}
In order to compute the constraints coming from the measurements of
$\Omega_\mathrm{HI}$, we use a second Fisher matrix 
\begin{equation}
    G_{ij} = \frac{1}{\sigma^2_{\Omega_\mathrm{HI}}}\,\dfrac{\partial \Omega_\mathrm{HI}}{\partial \theta_i}  \dfrac{\partial\Omega_\mathrm{HI} }{\partial \theta_j}\;,
    \label{eqn:fisher_density}
\end{equation}
with $\sigma^2_{\Omega_\mathrm{HI}}$ the standard error of the
(observed) HI cosmic abundance.
The combined constraints from the MFs and $\Omega_\mathrm{HI}$
are obtained using $(\mathsf{F}+\mathsf{G})^{-1}$.

\section{Results} \label{sec:results}
In this section, we present the main results of our study:
a Fisher forecast for the HIHMR. As a reference, we consider a (future) wide HI intensity-mapping survey conducted with the SKA-1 MID observatory. The experiment was originally designed in order to
measure the power spectrum of the maps and subsequently
constrain the dark-energy equation of state at a competitive level with respect to the forthcoming generation of galaxy redshift surveys \citep{bull2015}.
\subsection{The SKA-1 MID dark-energy survey}
\label{sec:SKA-1MID}

This intensity-mapping survey is expected to cover an area of the sky
of $\Omega_\mathrm{surv}\simeq 20,000$ deg$^2$ (corresponding to $\zeta\approx 5.8$)
and a frequency range of
$350<\nu<1050$ MHz (i.e. $0.35 <z< 3.06$).
The total observation time should be of approximately 10,000 h. 
Using the same pixel size as in section~\ref{sec:instnoise}, this set-up corresponds to $t_\mathrm{pix}=9.5\times 10^4$ s which should
be enough to determine the HIHMR in a nearly optimal way.

\subsection{Constraints from individual channels}
\label{sec:onechannel}

\begin{figure}
\includegraphics[width=1\columnwidth]{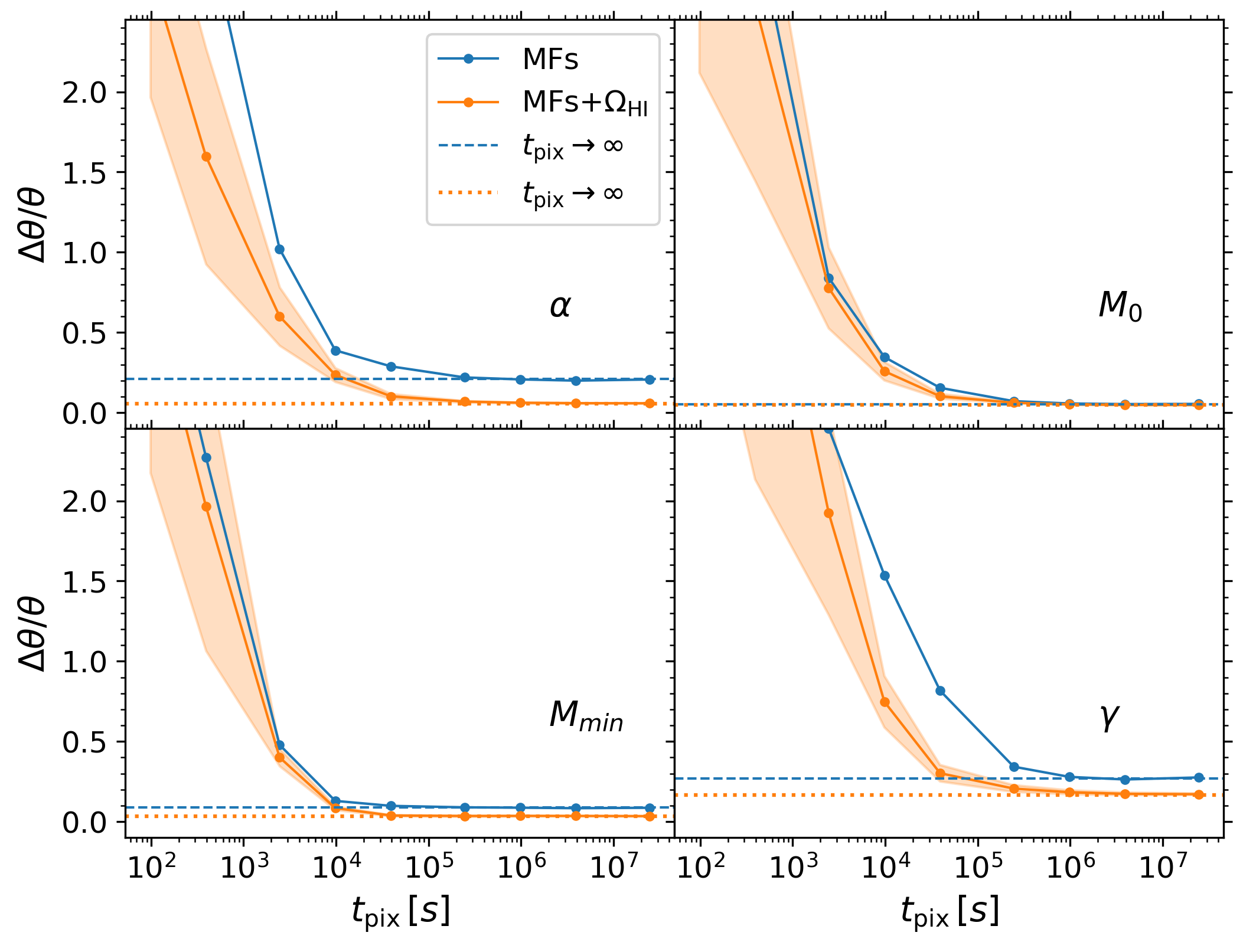} 
\caption{The relative marginalised uncertainty (68.3 per cent credibility interval) for each model parameter as a function of  $t_\mathrm{pix}$. The orange lines and shaded regions show the mean
and rms scatter of the results over the 30 noise realisations.
They refer to the combination of the MFs with the $\Omega_\mathrm{HI}$
data.
The blue lines show the mean uncertainty when only the MFs are considered (in order to improve readability, we do not show the scatter which is similar to the previous case).
Finally, the dotted and dashed lines indicate the results obtained in the absence of thermal noise, i.e. in the limit  $t_\mathrm{pix} \to \infty$.
}

\label{fig:constraints-tpix}
\end{figure}

Our Fisher forecasts are presented in figure~\ref{fig:constraints-tpix}. We first discuss the ideal situation corresponding to
$t_\mathrm{pix} \to \infty$ which we use as a reference and represent with horizontal lines in the figure). By combining measurements of the MFs and $\Omega_\mathrm{HI}$,
all parameters of the mean HIHMR are constrained to better than 10 per cent (68.3 per cent credibility) as indicated by the orange dotted lines.
Of course, these constraints progressively deteriorate by decreasing $t_\mathrm{pix}$ (orange curves). However, they change very little
if $t_\mathrm{pix}\gtrsim 4\times 10^4$ s
and worsen significantly only if $t_\mathrm{pix}\ll 10^4$ s.

Considering only the MFs (blue curves) deteriorates the constraints on $\alpha$ and $\gamma$ by nearly a factor of 2 while it does not affect the other parameters for large values of $t_\mathrm{pix}$ (blue dashed lines).
On the other hand, for lower $t_\mathrm{pix}$, the constraints
on all parameters become worse.

\begin{figure}
\includegraphics[width=1.\columnwidth]{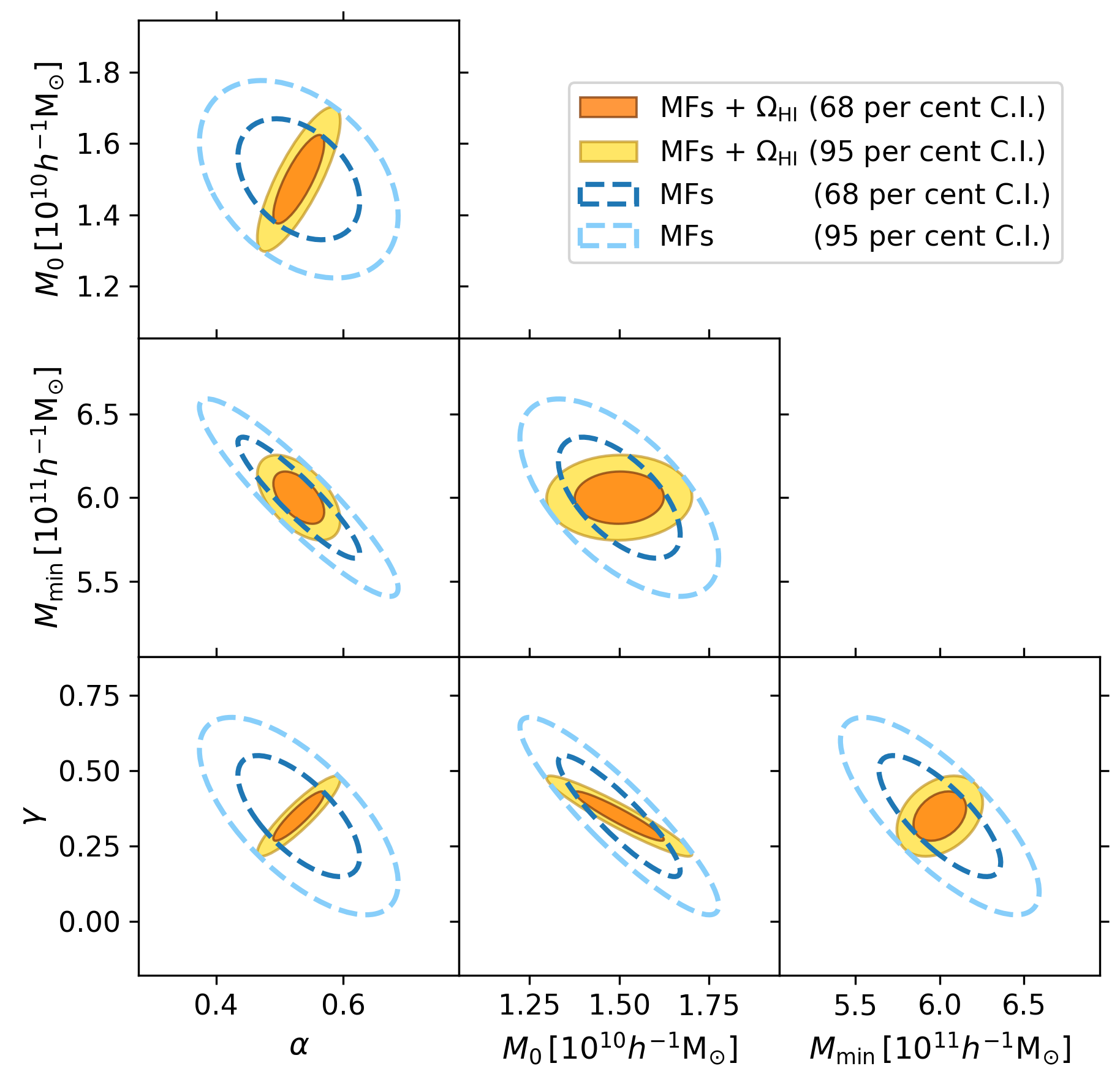} 
\caption{Joint marginalised $68.3$ per cent and $95.4$ per cent credible regions for all pairs of model parameters that determine the
HIHMR. The figure refers to a particular realisation of thermal noise obtained assuming $t_\mathrm{pix} = 3.9 \times 10^4$ s. Dashed ellipses show the results obtained from the Fisher information matrix for the MFs while shaded ellipses refer to the combination with measurements of $\Omega_\mathrm{HI}$.}

\label{fig:fisher}
\end{figure}

An example of the joint (68.3 and 95.4 per cent) credible regions for all parameter pairs is shown in figure~\ref{fig:fisher}. 
This plot has been obtained from one particular realization 
of thermal noise with $t_\mathrm{pix} = 3.9\times 10^4$ s. 
The dashed lines show the results obtained from the MFs while
the shaded regions refer to the combination with $\Omega_\mathrm{HI}$. 

\begin{figure}
\includegraphics[width=1\columnwidth]{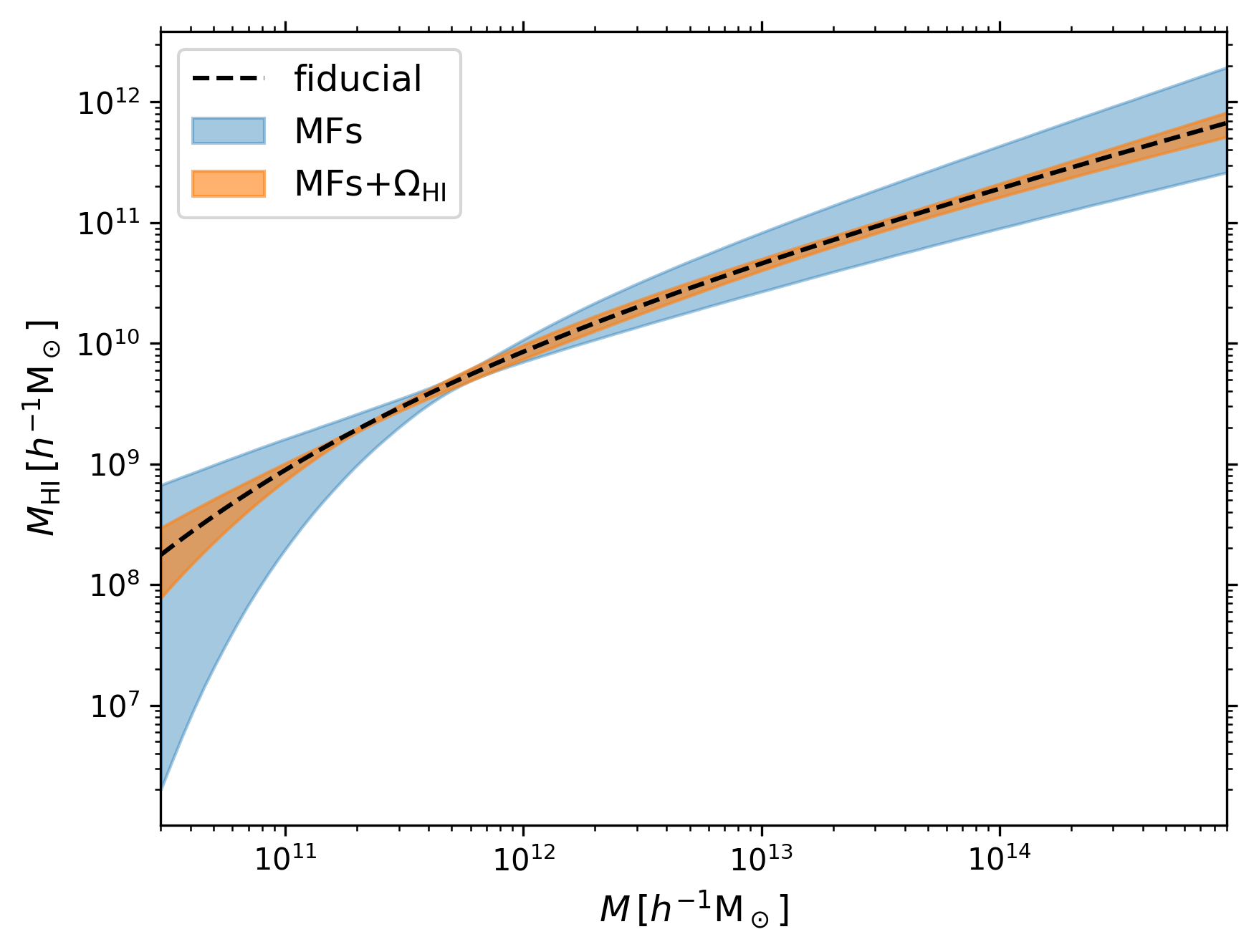} 
\caption{Central 68.3 per cent credibility interval of the mean HIHMR at $z=1$ obtained by fitting the MFs (light blue) and their combination with estimates of $\Omega_\mathrm{HI}$ (orange) in a single frequency channel at $z=1$. 
This result is obtained by marginalising 
equation~(\ref{eqn:MHI_Mh}) over
the posterior distribution of the model parameters shown in figure~\ref{fig:fisher}, and thus assumes $t_\mathrm{pix} = 3.9 \times 10^4$ s. 
The dashed curve indicates the fiducial relation.
}
\label{fig:posterior}
\end{figure}

In figure~\ref{fig:posterior}, we show that the MFs and $\Omega_\mathrm{HI}$
are very informative regarding the halo occupation properties of HI.
The posterior distribution of the mean HIHMR shows little scatter around the fiducial model.
Uncertainties on $M_{\mathrm{HI}}$ are particularly small for halo masses $M\simeq 10^{12}\,h^{-1}$ M$_\odot$ (which corresponds to a few $M_\mathrm{min}$ and identifies the haloes containing most HI) and increase as one moves away in both directions.

\begin{table*}
    \caption{Fiducial values and forecast uncertainties for the model parameters that regulate the
    HIHMR. The top section gives the results from the Fisher-information analysis for one channel at $z=1$.
    The bottom sections refer to fits of either second-order polynomials, 
    $\theta(z) = \theta_0 + \theta_1\,(z-1) + \theta_2\,(z-1)^2$ or
    constants, $\theta(z)=\theta_0$, to the measurements in the data cubes for $0.5<z<1.5$. The optimistic case (opt) treats all channels as independent data points while the pessimistic one (pes) considers only one channel every five.}
    \label{tab:forecasts}
    \centering
    \begin{tabular}{ccccccccccccccc}
    \hline
 & \multicolumn{3}{c}{$\alpha$} & & 
 \multicolumn{3}{c}{$M_0$} & &
 \multicolumn{3}{c}{$M_{\rm min}$} & &$\gamma$\\
 & \multicolumn{3}{c}{}& &\multicolumn{3}{c}{$[10^{10} h^{-1}\,\mathrm{M}_\odot]$} && \multicolumn{3}{c}{$[10^{11} h^{-1}\,\mathrm{M}_\odot]$} && & \\ \hline \hline
 Fiducial value $\theta$ & 0.53 & & & & 1.50 & & & & 6.00 & & & & 0.35  \\
$\Delta \theta$ single channel &0.05 & & & & 0.15& & & & 0.23& & && 0.11\\
\hline
& $\theta_0$ & $\theta_1$ & $\theta_2$ & & $\theta_0$ & $\theta_1$ & $\theta_2$ & & $\theta_0$ & $\theta_1$ & $\theta_2$ & & $\theta_0$  \\
\hline
Fiducial value $\theta_i$ & 0.53 & 0.20 & -0.06 && 1.50 & -1.49 & 1.38 && 6.00 & -8.21 & 6.02 && 0.35 \\
$\Delta \theta_i$ data cube (opt) & 0.01 & 0.01 & 0.06 & & 0.02 & 0.05 & 0.16 & & 0.02 & 0.07 & 0.23& & 0.01\\
$\Delta \theta_i$ data cube (pes) & 0.01 & 0.03 & 0.11 & & 0.03 & 0.09 & 0.33 & & 0.05 & 0.15 & 0.49 & & 0.02\\
 \hline
    \end{tabular}
\end{table*}

\subsection{Constraints from the full data cube}
\label{sec:multichannel}
So far, we have focused on a single frequency channel corresponding to a narrow redshift bin centered around $z = 1$ and constrained
the HIHMR
using the  MFs of the two-dimensional HI intensity map.
We want now to make use of the full data cube that the radio observations will provide. With this aim in mind, we imagine to repeat the measurement of the MFs in each frequency channel and then fit 
the redshift evolution of the model parameters. 
In fact, it is reasonable to expect that the parameters of the HIHMR should vary smoothly with redshift. 
In order to exemplify the power of this technique,
we synthesize the data at $z\simeq 1$ in table~1 of \citet{21cmIM} in terms of the second-order Taylor expansions
\begin{align}
    \alpha&=0.53+0.18\,(z-1)-0.11\,(z-1)^2\;,\\
    M_0&=[1.5-1.5\,(z-1)+1.3\,(z-1)^2]\times 10^{10}\,h^{-1}\,\mathrm{M}_\odot\;,\\
    M_\mathrm{min}&=[6.0-8.2\,(z-1)+5.8\,(z-1)^2]\times 10^{11}\,h^{-1}\,\mathrm{M}_\odot\;, \\
    \gamma&=0.35\;, 
\end{align}
and use them as the fiducial model for the evolution of the
HIHMR in the redshift range $0.5<z<1.5$.
This interval contains 189 channels with $\Delta \nu=2$ MHz.
We can measure the MFs in each of them and then fit the marginalised
constraints on the individual model parameters with second-order polynomials of $z-1$. Results obtained by assuming independent data points are reported in table~\ref{tab:forecasts}.
Since the thickness of each channel along the line of sight is only a few times larger than the correlation length of the density field and redshift-space distortions can also shift HI by comparable distances, it might be inappropriate to consider the results obtained from consecutive channels as independent.  
We thus repeat the analysis by considering only one channel every five. Of course, the uncertainty on the model parameters slightly increases in this case (see table~\ref{tab:forecasts}).
Note that, if we compare the results obtained for $z=1$ from
the multi-channel fits to those presented in section~\ref{sec:onechannel},
we note an improvement of the error bars
up to a factor of ten.


\section{Discussion}
\label{sec:discussion}
It has been shown that the MFs are powerful data-analysis tools to characterise --and extract information from--
various cosmological datasets.
However, they are sub-optimal 
in the case of Gaussian random fields for which the whole statistical information is contained in the power spectrum.
In fact, many of the practical issues associated with the MFs (dealing with a masked sky, modelling the impact of noise) are solved problems for inferences based on power spectra.
A question then naturally arises: are the fluctuations in the 21-cm-brightness-temperature distribution departing from a Gaussian
random field so that to justify the use of the MFs in our study? 
This question is particularly relevant given the large beam of the SKA-1 MID telescopes which corresponds to a transverse size of approximately 50 comoving Mpc at $z=1$ (that we use in combination with frequency channels that extend for nearly 13 comoving Mpc
along the line of sight). 

\begin{figure}
\includegraphics[width=1\columnwidth]{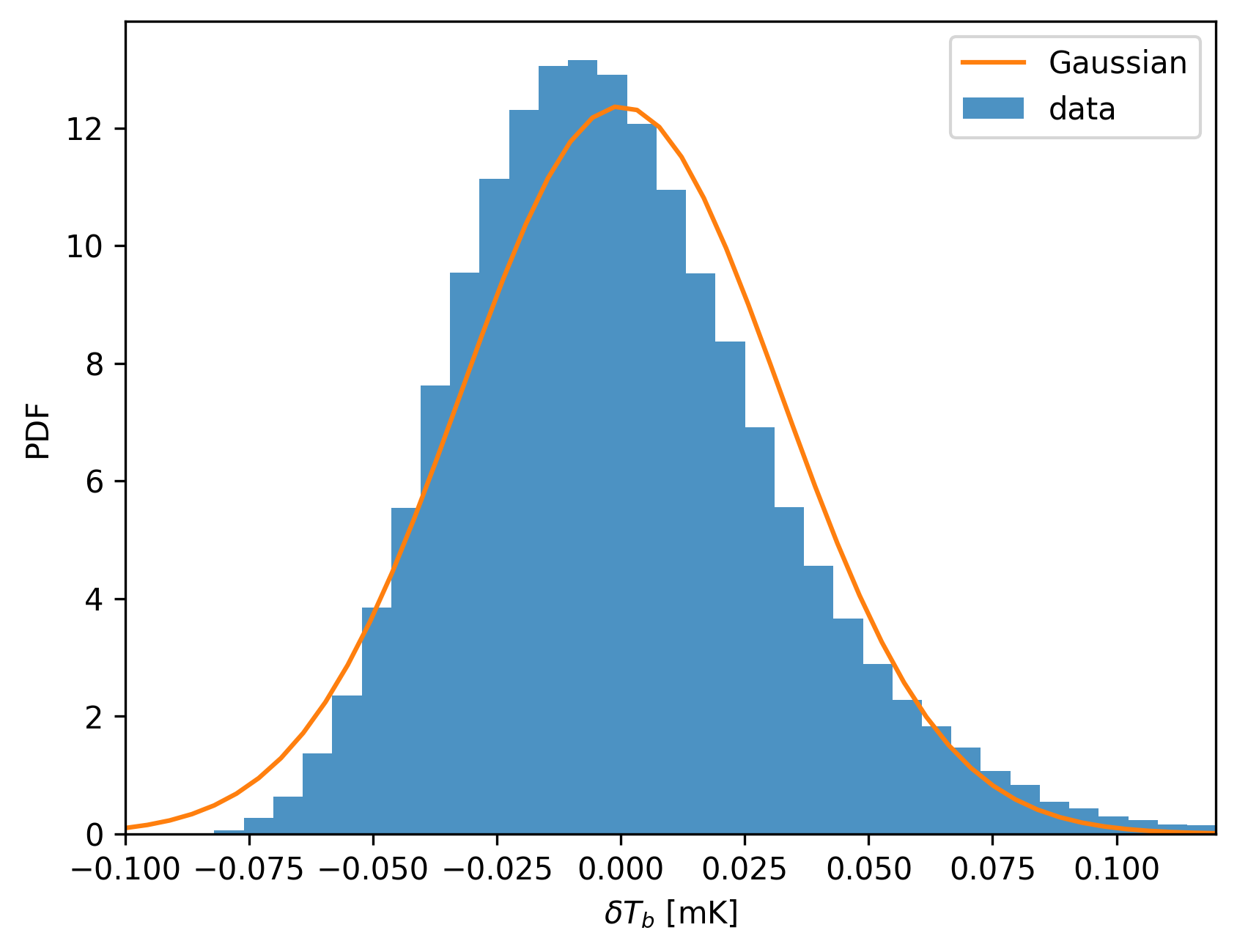} 
\caption{
The PDF of
the brightness-temperature fluctuations in
our mock maps (shaded histogram) is contrasted with a Gaussian distribution with the same mean and variance.}

\label{fig:PDFgaussian}
\end{figure}

In figure \ref{fig:PDFgaussian}, we show the probability density function (PDF) of the brightness temperature extracted from our simulation box in the absence of thermal noise (i.e. assuming $t_\mathrm{pix} \rightarrow \infty$). 
In order to facilitate
visual comparison with the Gaussian case,
we also plot a normal distribution with the same mean and standard deviation as
the simulated data (solid line). 
The PDF of the brightness temperature is
clearly asymmetric (with a skewness
of $1.04\pm 0.01$)
and has heavier tails 
than a Gaussian distribution 
(with a kurtosis of $0.708\pm0.006$)

\begin{figure}
\includegraphics[width=1\columnwidth]{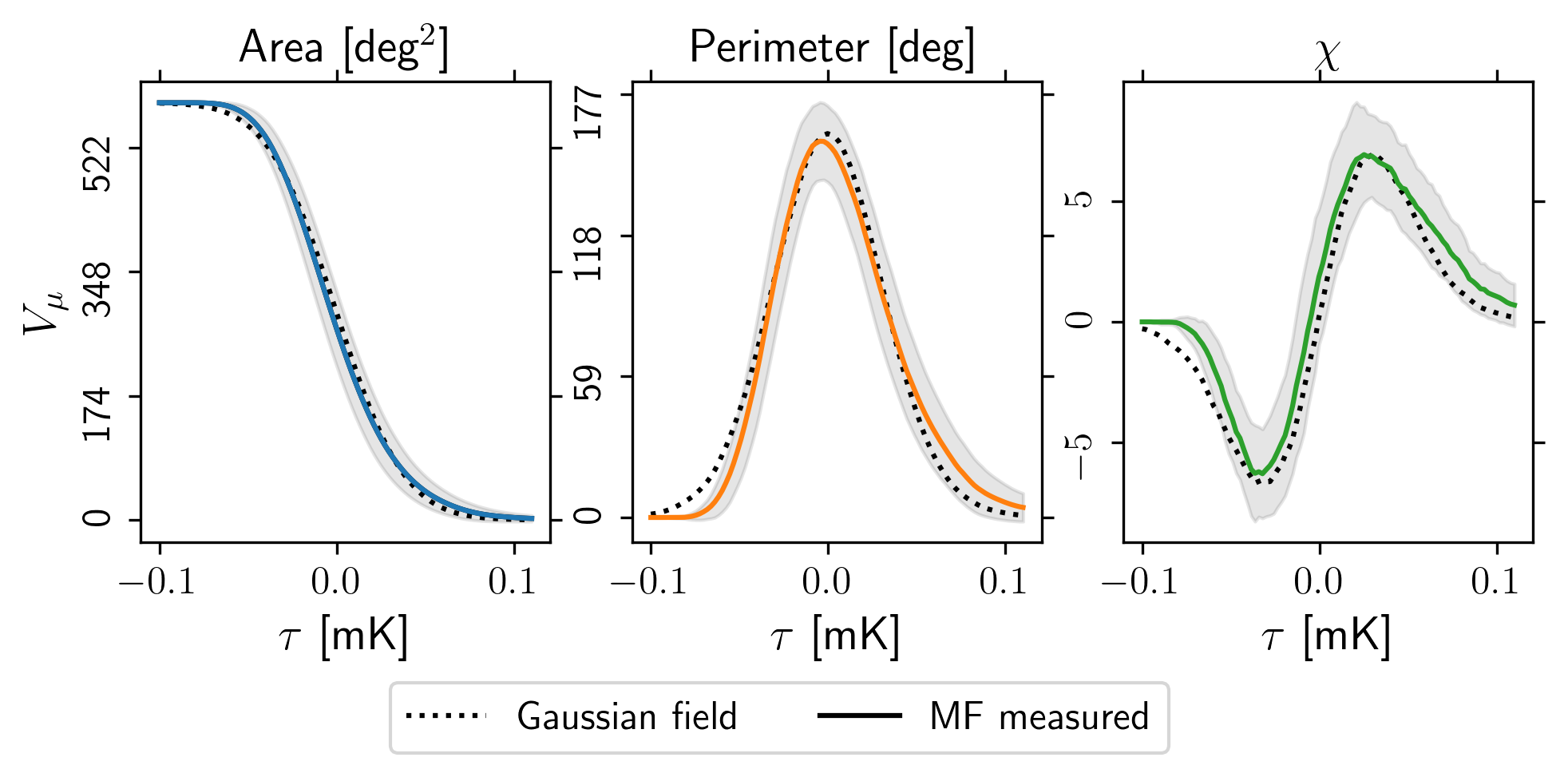} 
\caption{The mean MFs (solid) and their rms scatter (shaded region) extracted from our simulations in the absence of thermal noise
are compared with the mean signal
for a Gaussian random field with the same
power spectrum and domain (dotted lines).}
\label{fig:MFgaussian}
\end{figure}

It is also interesting to check how much the MFs in the mock maps depart from 
the expected values for a Gaussian random field. Analytical formulae have been
derived for the MFs of a Gaussian random field on the Euclidean plane or on the two sphere (see section \ref{sec:fisher_minkowski}).
However, important corrections need to be applied\footnote{This is often neglected in
astrophysical and cosmological applications but see \cite{Pranav+19} for a nice example.}
when the domain of the random field is a small subset of the above mentioned spaces. Moreover, the overall normalisation
of the curves $V_i(\tau)$ depends on the power spectrum of the random field.
For these two reasons, we compute the Gaussian MFs by
averaging over many realisations obtained
by shuffling the phases between the Fourier modes of the simulated data (in the absence of thermal noise, as in figure~\ref{fig:MFvar}).
Our results are shown 
in figure~\ref{fig:MFgaussian}. 
While the surface area $V_0(\tau)$ covered by the excursion set is hardly distinguishable from the Gaussian case,
larger differences are noticeable 
for the perimeter of the boundary $V_1(\tau)$ and
for the Euler characteristic,
in particular, for extreme values of the
threshold parameter $\tau$ both on the low and high sides.

A related issue is whether the information encoded in the MFs on the large scales probed by the SKA-1 MID is enough
to constrain all the parameters that
influence the mean HIHMR. In fact, by reasoning in terms of the halo model \citep[see][for a review]{Cooray-Sheth02}, 
one expects the data to be fully in the 
so-called `two-halo' regime where all the information about the HIHMR is collapsed into the linear bias parameter of the HI with respect to the mass.
Effectively, this would mean that the exercise we are proposing tries to constrain several parameters from the measurement of a single quantity. 
This line of reasoning assumes that the bias relation between HI and matter fluctuations is linear on scales of 10-50 Mpc. 
It is well known, however, that this is only true to first approximation. Several non-linear corrections are needed to accurately model the spatial distribution of dark-matter haloes, especially when one
considers statistics that are more sensitive to non-Gaussian features than
the power spectrum. State-of-the-art models include, at least, extra terms that scale quadratically with the matter density or linearly with the tidal field \citep[see e.g.][for a review]{Desjacques+18}. Our results can thus be interpreted as providing evidence that the same corrections are needed to describe the non-Gaussian features of the HI distribution.
Further evidence has been recently provided by \citet{Cunnington+21} in a study of the bispectrum of 21-cm intensity maps.
Nevertheless, some level of degeneracy between the model parameters is indeed present in our results, as shown in figure~\ref{fig:fisher}. This is mostly broken when independent measurements of the HI abundance are considered in combination with the MFs.

\section{summary}
\label{sec:conclusion}
Determining the overall content and the spatial distribution of HI in the post-reionization Universe is pivotal to understand galaxy formation and evolution.
An important step in this direction is the determination of the HIHMR which gives 
the mean and scatter of the total HI mass contained within a dark-matter halo of mass $M$.
In this paper, we have investigated the possibility of constraining parametric models of the HIHMR from 21 cm intensity maps at redshift $z\simeq 1$.
In particular, we have used the
geometry and topology of the maps as quantified by the MFs of the
brightness-temperature isocontours.

For practical reasons, we have considered a specific parameterization for the HIHMR in which the mean HI mass at fixed
halo mass is given by equation~(\ref{eqn:MHI_Mh}) and a lognormal scatter of size $\sigma$ is assumed (see section~\ref{sec:relation}).
By assuming a set of fiducial values based on previous numerical studies,
we have generated mock data from a large N-body simulation and used the Fisher information matrix to derive the forecast constraints on the model parameters. 
As a reference case, we have considered the SKA-1 MID dark-energy survey at $z=1$ (conducted in single-dish mode) and assumed frequency channels of width $\Delta \nu=2$ MHz.

Our main results can be summarised as follows.
\setcounter{enumi}{2}
\renewcommand{\theenumii}{\roman{enumii}}
\begin{enumerate}
\item After subtracting the foregrounds, the 21~cm signal is
of the order of a few mK, nearly 1000 times lower than the system temperature of the telescopes. In order to beat thermal noise and reach the sensitivity necessary for imaging the HI distribution, long integration times per pixel, $t_\mathrm{pix}$, are thus required. By using the MFs of the 21~cm intensity map in a single frequency channel, we find that the parameters of the HIHMR can be measured with a signal-to-noise ratio of one if $t_\mathrm{pix}\simeq 9.8\times 10^3$ s. 
Nearly optimal error bars are obtained using $t_\mathrm{pix}\simeq$ a few $\times 10^4$ s.
This corresponds to a total observing time of 4 - 5 days.
\item 
Information on the mean HI density, $\Omega_\mathrm{HI}$, is lost during the foreground removal from the intensity maps. This loss can be restored by combining the MFs with independent measures of $\Omega_\mathrm{HI}$. 
This addition slightly improves the constraints on some of the parameters that regulate the HIHMR, in particular on the slope $\alpha$ defined in equation~(\ref{eqn:MHI_Mh}). 
\item The mean HIHMR is very tightly constrained for haloes with $M\simeq 10^{12} h^{-1}$ M$_\odot$ which contain most of the HI at $z=1$. Uncertainties grow larger as both larger and smaller halo masses are considered.
\item Combining measurements of the MFs in different frequency channels provides exquisite constraints on the redshift evolution of the HIHMR (both for the mean and the scatter), especially for redshifts that lie around the center of the data cube along the frequency axis. In this case, we forecast uncertainties on the model parameters that are an order of magnitude smaller with respect to those extracted from a single channel.

\end{enumerate}

In this paper, we have used the SKA-1 MID dark-energy survey as an example of what can be achieved with forthcoming facilities.
Observations at higher angular resolution with SKA precursors like the Hydrogen Intensity and Real-time Analysis eXperiment \citep[HIRAX,][]{hirax} in the southern hemisphere and the Canadian Hydrogen Intensity Mapping Experiment\footnote{As well as other proposed facilities such as the Canadian Hydrogen Observatory and Radio-transient Detector \citep[CHORD,][]{Vanderlinde+19}, the Deep Synoptic Array-2000 \citep[DSA-2000,][]{Hallinan+19}, and the Packed Ultra-wideband Mapping Array \citep[PUMA,][]{Slosar+19}.} \citep[CHIME,][]{Bandura+14} in the northern hemisphere will probe length scales at which the HI-to-mass bias relation is more non-linear. Given long enough integration times, these experiments might be able to partially remove the degeneracy among the model parameters of the HIHMR and even constrain them more tightly. We will investigate this possibility in our future work.

\section*{Acknowledgements}
We are grateful to an anonymous reviewer for drawing our attention to the topics discussed in section~\ref{sec:discussion}.
We thank Moutaz Haq for making the \textsc{minkfncts2d} code public as well as Sabino Matarrese, Alkistis Pourtsidou, Marta Spinelli and Matteo Viel for discussions during the early stages of this work. 
This work is carried out within the Collaborative Research Centre 956 ‘The Conditions and Impact of Star Formation’, sub-project C4, funded by the Deutsche Forschungsgemeinschaft (DFG).
CS aknowledges the Programme National Cosmologie et Galaxies (PNCG) of CNRS/INSU with INP and IN2P3 for financial support.
The CosmoSim database used in this paper is a service by the Leibniz-Institute for Astrophysics Potsdam (AIP). The MultiDark database was developed in cooperation with the Spanish MultiDark Consolider Project CSD2009-00064.

\section*{Data availability statement}
The data underlying this article will be shared on reasonable request to the corresponding author.

\bibliographystyle{mnras}
\bibliography{main}

\bsp
\label{lastpage}
\end{document}